\RequirePackage{fix-cm}
\documentclass[twocolumn,epjc3]{svjour3}  
\smartqed  % flush right qed marks, e.g. at end of proof
\RequirePackage{graphicx}
\RequirePackage{xcolor}
\RequirePackage{url}
\RequirePackage{soul}
\RequirePackage{caption}
\RequirePackage{subcaption}
\RequirePackage{array}
\RequirePackage{tabularx}
\RequirePackage{textcomp}
\RequirePackage{upgreek}

\RequirePackage[numbers,sort&compress]{natbib}
\RequirePackage[colorlinks,citecolor=blue,urlcolor=blue,linkcolor=blue]{hyperref}
\journalname{Eur. Phys. J. A}
\begin{document}

\title{Silicon tracker array for RIB experiments at SAMURAI%\thanksref{t1}
}


\author{A.I. Stefanescu\thanksref{addr1,addr2,addr3}
        \and
        V. Panin\thanksref{addr3,addr4} %etc.
        \and
        L. Trache\thanksref{e1,addr1}
        \and
        T. Motobayashi\thanksref{addr3}
        \and
        H. Otsu\thanksref{addr3}
        \and
        A. Saastamoinen\thanksref{addr5}
        \and
        T. Uesaka\thanksref{addr3}
        \and
        L. Stuhl\thanksref{addr6}
      \and
        J. Tanaka\thanksref{addr3}  
       \and
        D. Tudor\thanksref{addr1,addr2,addr3} 
        \and
        I.C. Stefanescu\thanksref{addr1,addr2}
        \and
        A.E. Spiridon\thanksref{addr1}
        \and
        K. Yoneda\thanksref{addr3}
        \and
        H. Baba\thanksref{addr3}
        \and
        M. Kurokawa\thanksref{addr3}
        \and
        Y. Togano\thanksref{addr7}
        \and
        Z. Halasz\thanksref{addr8}
        \and
        M. Sasano\thanksref{addr3}
        \and
        S. Ota\thanksref{addr6}
        \and
        Y. Kubota\thanksref{addr3}
        \and
        D.S. Ahn\thanksref{addr3}
        \and
        T. Kobayashi\thanksref{addr9}
        \and
        Z. Elekes\thanksref{addr8}
        \and
        N. Fukuda\thanksref{addr3}
        \and
        H. Takeda\thanksref{addr3}
        \and
        D. Kim\thanksref{addr3,addr10}
        \and
        E. Takada\thanksref{addr11}
        \and
        H. Suzuki\thanksref{addr3}
        \and
        K. Yoshida\thanksref{addr3}
        \and
        Y. Shimizu\thanksref{addr3}
        \and
        H.N. Liu\thanksref{addr12}
        \and
        Y.L. Sun\thanksref{addr12}
        \and
        T. Isobe\thanksref{addr3}
        \and
        J. Gibelin\thanksref{addr13}
        \and
        P.J. Li\thanksref{addr14}
        \and
        J. Zenihiro\thanksref{addr3}
        \and
        F.M. Marqu\'es\thanksref{addr13}
        \and
        M.N. Harakeh\thanksref{addr15}
        \and
        G.G. Kiss\thanksref{addr8}
        \and
        A. Kurihara\thanksref{addr16}
        \and
        M. Yasuda\thanksref{addr16}
        \and
        T. Nakamura\thanksref{addr16}
        \and
        S. Park\thanksref{addr10}
        \and
        Z. Yang\thanksref{addr3}
        \and
        T. Harada\thanksref{addr17}
        \and
        M. Nishimura\thanksref{addr3}
        \and
        H. Sato\thanksref{addr3}
        \and
        I.S. Hahn\thanksref{addr10}
        \and
        K.Y. Chae\thanksref{addr18}
        \and
        J.M. Elson\thanksref{addr19}
        \and
        L.G. Sobotka\thanksref{addr19}
        \and
        and
        C.A. Bertulani\thanksref{addr20}
}

\thankstext{e1}{e-mail: livius.trache@nipne.ro}
%\thankstext{e2}{e-mail: alexandra.chilug@nipne.ro}

\institute{Horia Hulubei National Institute for R$\&$D in Physics and Nuclear Engineering, IFIN-HH, 077125 București-Măgurele, Romania \label{addr1}
           \and
           Doctoral School of Physics, University of Bucharest, 077125 București-Măgurele, Romania \label{addr2}
           \and
           GSI Helmholtzzentrum für Schwerionenforschung GmbH, 64291, Darmstadt, Germany \label{addr3}
           \and
           RIKEN Nishina Center for Accelerator-Based Science, 2-1 Hirosawa, Wako, Saitama 351-0198, Japan \label{addr4}
           \and
           Cyclotron Institute, Texas A\&M University, College Station, TX-77843, USA \label{addr5}
            \and
           Center for Nuclear Study, University of Tokyo, 2-1 Hirosawa, Wako, Saitama, Japan \label{addr6}
           \and
           Department of Physics, Rikkyo University, Tokyo 171-8501, Japan \label{addr7}
           \and
           Atomki, Eötvös Lor\'and Research Network (ELKH), Debrecen, Hungary \label{addr8}
           \and
           Department of Physics, Tohoku University, Miyagi 980-8578, Japan \label{addr9}
           \and
           Department of Physics, Ewha Womans University, 120-750 Seoul, Korea \label{addr10}
           \and
           National Institute of Radiological Sciences (NIRS),4-9-1 Anagawa, Inage, Chiba 263-0024, Japan \label{addr11}
           \and
           D\'{e}partement de Physique Nucl\'{e}aire, IRFU, CEA, Universit\'{e} Paris-Saclay, F-91191 Gif-sur-Yvette, France \label{addr12}
           \and
           LPC CAEN, ENSICAEN, 6 bd Marchal Juin, 14050 Caen, Cedex, France \label{addr13}
           \and
           Department of Physics, The University of Hong Kong, Hong Kong, China \label{addr14}
           \and
           Nuclear Energy Group, ESRIG, University of Groningen, Zernikelaan 25, 9747 AA, Groningen, The Netherlands \label{addr15}
           \and
           Department of Physics, Tokyo Institute of Technology, 2-12-1 O-Okayama, Meguro, Tokyo 152-8551, Japan \label{addr16}
           \and
           Department of Physics, Toho University, 5-21-16 Omorinishi, Ota, 143-8540 Tokyo, Japan \label{addr17}
           \and
           Department of Physics, Sungkyunkwan University, Suwon 16419, Korea \label{addr18}
           \and
           Departments of Chemistry and Physics, Washington University, St. Louis, Missouri 63130, USA \label{addr19}
           \and
            Department of Physics and Astronomy, Texas A\&M University-Commerce, Commerce, TX, USA\label{addr20}
}

\date{Received: date / Accepted: date}

\maketitle

\begin{abstract}
This work describes a silicon tracker system developed for experiments with proton-rich radioactive ion beams at the SAMURAI superconducting spectrometer of RIBF at RIKEN. The system is designed for accurate angular reconstruction and atomic number identification of relativistic heavy ions and protons which are simultaneously produced in reactions motivated by studies of proton capture reactions of interest for nuclear astrophysics. The technical characteristics of the tracking array are described in detail as are its performance in two pilot experiments. The physics justification for such a system is also presented.

\end{abstract}

\section{Introduction}
\label{intro}
A principal aim of nuclear physics experiments with radioactive ion beams (RIB) at intermediate-high energies is to elucidate the structure and interaction of exotic nuclei far away from the $\beta$-stability line, on both the proton- or neutron-rich sides of the nuclear chart. 
While done at high (often relativistic) energy, the data can be used to evaluate the physics quantities relevant for low-energy astrophysics. This is accomplished by using indirect methods like: nuclear breakup, Coulomb dissociation, Trojan Horse Method, or asymptotic normalization coefficients (ANC). Some of these methods were successfully applied in previous works detailed in Refs. \cite{Trache2001,Banu2012,Motobayashi94,Tumino2018,Tribble}.
The study of the dissociation at intermediate energies in a reaction 
\(X \rightarrow Y + p\) can be used to evaluate the astrophysical S-factor and/or the reaction rate for the radiative proton-capture \(Y(p,\upgamma)X\). One of the case studies of this paper is the breakup of $^{9}$C in nuclear and Coulomb fields, a reaction used to extract the astrophysical S-factor, $S_{18}$ for the reaction \(^{8}B(p,\upgamma)^{9}C^{\ast}\) in hydrogen burning.

There is the need for highly-efficient exclusive measurements to compensate for the low intensities and poor quality and purity of secondary RIBs (as compared to primary-beam experiments). Due to the relativistic energies involved, reaction products are typically measured in strongly focused forward kinematics. This kinematics dictates the particle-detection and signal-processing requirements, e.g. fine granularity, high-density readout, and high rate capability.
To extract the relative energy as well as the scattering angle, tracking of the incoming and outgoing particles is essential in order to reconstruct the reaction kinematics.  Specifically, for accurate determination of the excitation energy of a decaying exit channel fragment, the relative angle between its decay particles must be measured with an excellent resolution. The method employed in the present effort is to measure relative angles before the particles enter a magnetic analyzer. This approach avoids the difficulties associated with the mixing of the direction and momentum information by the magnetic field. This strategy, however, has the complication that the two particles (often a proton and a heavy reaction product) move at similar angles in the forward direction, and cannot be measured in separate detectors. (This is not the case with measurements after passing through the magnetic field. On the other hand, the momenta of the exit channel fragments are measured in a good resolution by detecting them after passing through the magnetic analyzer. In this case entirely different detector systems were used to measure the decay particles.)

Our approach requires a detector system and pulse processing system with a huge dynamic range, so that both protons and heavy ions can be detected, as well as good position resolution. For example, in the case of the Coulomb dissociation studies, angular distribution measurements are essential to disentangle the contributions of E1 and E2 multipolarities to the total dissociation cross-section. That implies a good determination of both the direction of the incoming projectile on the target and the direction of the two, or more, emerging dissociation products. Another important experimental requirement is the momentum distribution measurement which offers information regarding the reaction mechanisms and single-particle structure, being the main experimental goal of the breakup studies. In brief, the detector system and the pulse-processing electronics must have a large dynamic range, be able to handle high counting rates, and be inexpensively replicated so that large channel numbers can be employed. The system must also be sufficiently versatile to handle the variety of RIB experiments envisioned. 

The detection array we report on here satisfies these requirements and is designed for the experiments with the SAMURAI spectrometer \cite{samurai} at the accelerator complex of the Nishina Center at RIKEN~\cite{nishina}. The system consists of an array of large-area silicon-strip detectors (SSD) equipped with front-end ASIC preamplifiers with dual-gain capabilities and external readout electronics based on the HINP technology~\cite{Engel2007}. This tracking system is designed to complement the experimental setup at SAMURAI, with an emphasis on the experiments with proton-rich RIBs. The SSD strip pitch assures position resolution at sub-mm level while the large dynamic range of the readout electronics ($\sim$ 3,000 - 10,000) allows for coincident measurements of protons and heavy ions produced in the same reaction. The technical specifications of the silicon detectors are sketched in Sect. 2, Sect. 3 gives detailed description of the front-end electronics, Sect. 4 shows the performance of the described system as observed in test experiments, and our conclusions are presented in Sect. 5.

\section{Silicon detectors system}

High granularity of the silicon detectors is the key ingredient to resolve different particle tracks and to obtain position information with high resolution. This is ensured by choosing Si-strip detectors with a small pitch. We chose GLAST-type \cite{Bellazzini512,Ohsugi541} 325 {\textmu}m thick silicon-strip detectors (SSD) made by Hamamatsu Photonics Company. These single-sided detectors are made of bulk n-doped silicon with p$^{+}$ DC-coupled implant strips, built on 6" high-resistivity wafers having an active area of 87.5 $\times$ 87.5 mm$^2$.  The width of the silicon implant strip is 56  {\textmu}m and the pitch is 228 {\textmu}m. 
The main technical characteristics of the GLAST sensors are listed in Table~\ref{table1} and the full technical description of these detectors can be found in Refs. \cite{Bellazzini512,Ohsugi541}.

\begin{table}[h!]
\begin{center}
\caption{Characteristics of the GLAST-type sensors.}\label{table1}
\begin{tabular}{@{}llll@{}}
\hline
Properties & Specifications \\
\hline
Active area    & 87.55 $\times$ 87.55 mm$^{2}$    \\
n-type substrate thickness    & 325 {\textmu}m    \\
Readout pitch size    & 228 $\times$ 3 = 684 {\textmu}m    \\
Final number of p$^{+}$ strips per layer    & 384 / 3 = 128    \\
Full depletion voltage    & -90 V   \\
Width of implant strip    & 56 {\textmu}m    \\
Interstrip capacitance & $<$ 7 pF \\
\hline
\end{tabular}
\end{center}
\end{table}

\begin{figure}[ht!]
 	\begin{center}
 		\includegraphics[width=0.8\columnwidth]{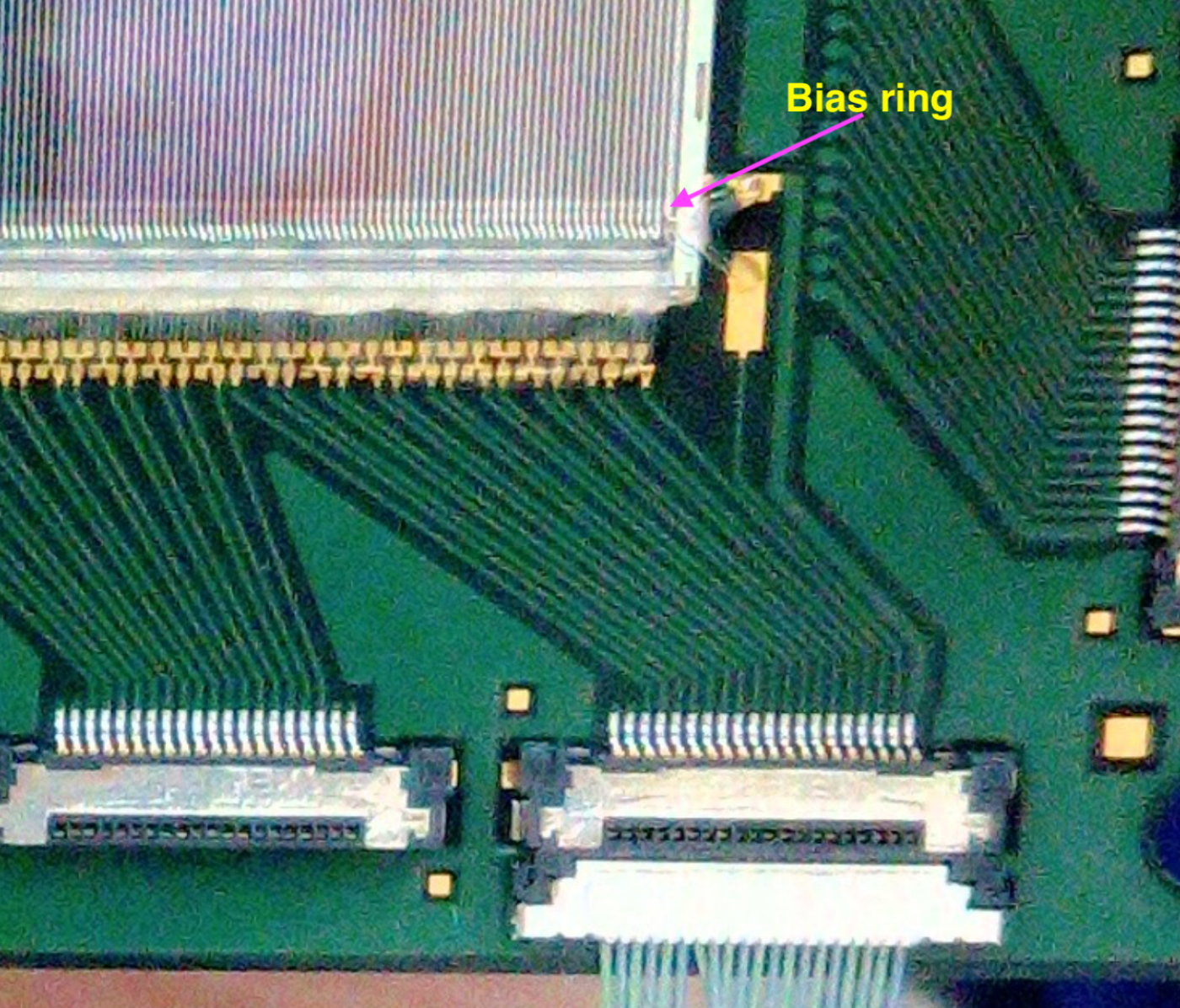}
 		\caption{\label{fig:connected_strips} Photo of the silicon-PCB showing that every 3 strips of the silicon layer are electrically coupled to one input of the dual-gain preamplifier (see text) resulting in the readout pitch of 684 {\textmu}m.}
 	\end{center}
 \end{figure}

In order to reduce the number of the readout channels without sacrificing the overall performance of the detection system, every three adjacent strips are electrically coupled (as shown in Fig.~\ref{fig:connected_strips}) resulting in a readout pitch size of 684 {\textmu}m and a final number of 128 readout strips per detector.
The read-out pitch of the silicon strips was set by compromise between physics performance and cost. There is little physics gain when the readout pitch becomes comparable with the multiple scattering effects in the target. Strip summing also reduces the effective inter-strip leakage, a problem for the heavy ions. In short, a smaller read-out pitch would increase the cost without improving the resolution of the desired experimental quantities. The design of the GLAST detector was chosen to minimize the strip cross talk by minimizing interstrip capacitance. This is accomplished by the narrow implant strip, the wide isolation strip (64 {\textmu}m), both of which are small compared to the actual pitch. In this Si design, each strip is extended underneath a poly-silicon resistor that is connected directly to the bias ring, for bias and isolation. In this way the sensitive area of the detector is maximized and the edge dead area is minimized. The second function of the bias ring is to absorb the leakage current outside the bias ring (Fig.~\ref{fig:connected_strips}). 

 In order to obtain the 2D (x,y) position, alternate (1D) silicon detectors are rotated by 90$^\circ$ and share the same PCB frame with their respective bondings soldered to each side of it, creating a back to back setup. The corresponding signals from the silicon strips are collected using ultra-fine coaxial cables with 0.4 mm pitch and the first and last connectors are used to distribute the bias voltage to the bias ring.
 
 \begin{figure}[h!]
 	\begin{center}
 		\includegraphics[width=\columnwidth]{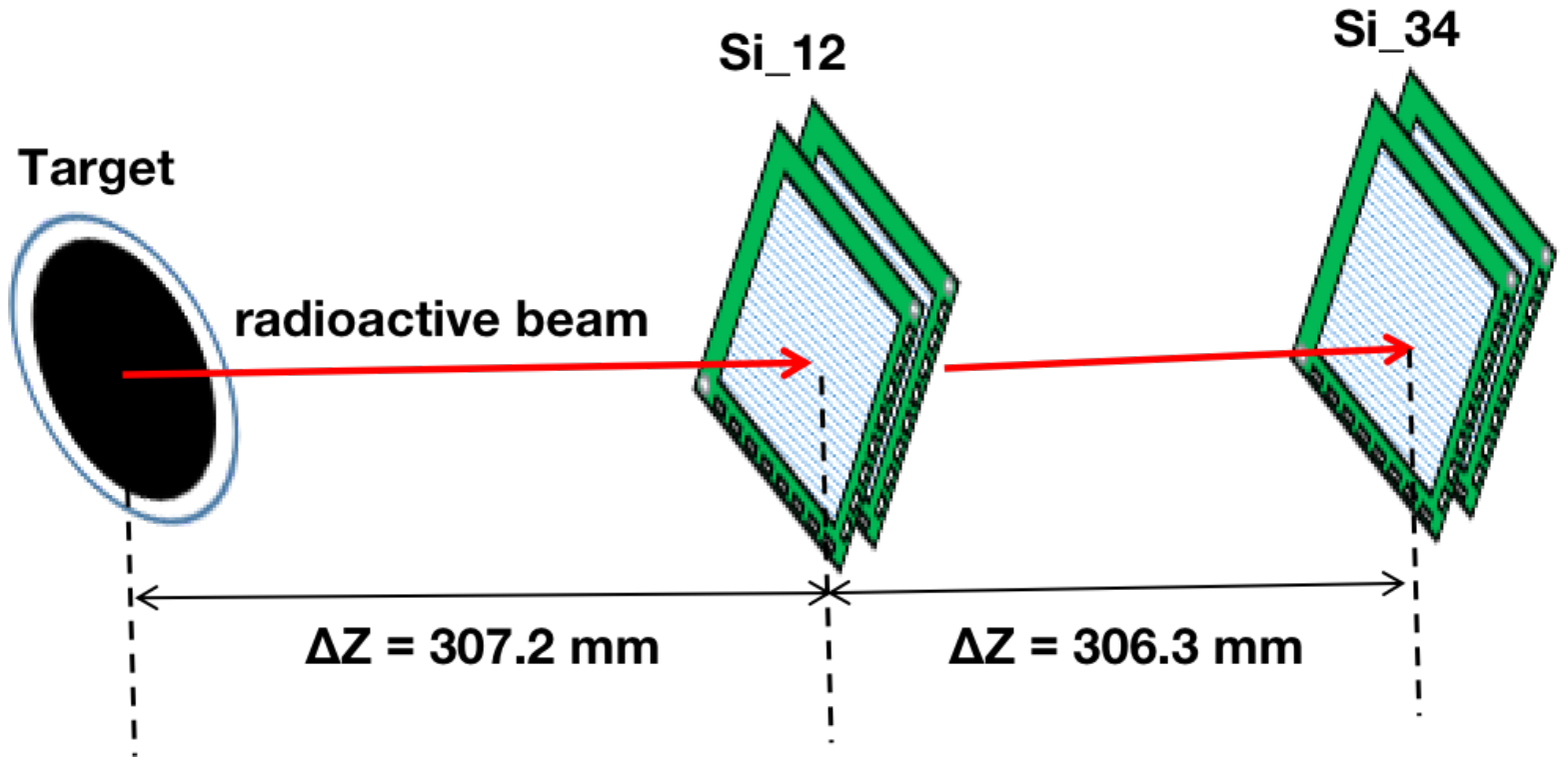}
 		\caption{\label{fig:geometry_experiment} Schematic arrangement of the silicon trackers during the experiments with the SAMURAI setup.}
 	\end{center}
 \end{figure}
 
The full tracking array consists of two SSD pairs arranged perpendicularly to the beam and centered on the beam axis (see Fig. \ref{fig:geometry_experiment}). For the purpose of mechanical and electrical compactness of the entire assembly (i.e. the use of shortest cables between the detectors and preamplifiers, and between preamplifiers and HINP, to reduce the noise and the delay and to fit the vacuum chamber in the experimental area together with the other SAMURAI detectors), every pair is rotated by 45$^\circ$ around the beam axis. The first pair is placed 30 cm downstream of the target and the distance between the two pairs is also 30 cm. The detectors are mounted into the beamline, between the reaction target and the entrance to the SAMURAI magnetic spectrometer in order to reconstruct the tracks of the emitted particles at small scattering angles around 0 degree. Additional detail can be found in Ref. \cite{chilug2019}.

\section{Front-End electronics}

The solution for the front-end electronics is dictated by the design requirement to instrument a large number of detector channels and by the dynamic range requirements. To solve the second issue, i.e. the system was designed to detect both fast protons and fragments up to Sn (Z = 50) with  kinetic energies between 100 -- 350 AMeV. This corresponds to the energy losses in 325 {\textmu}m silicon ranging from 100 -- 200 keV up to several hundred MeV.

Considering the number of signal channels from the four SSDs, it is necessary to use a highly integrated electronics employing application specific integrated circuit  (ASIC) technology designed so that the proton and the heavy-ion signals can be processed in coincidence. 

\begin{figure}[h!]
 	\begin{center}
 		\includegraphics[width=\columnwidth]{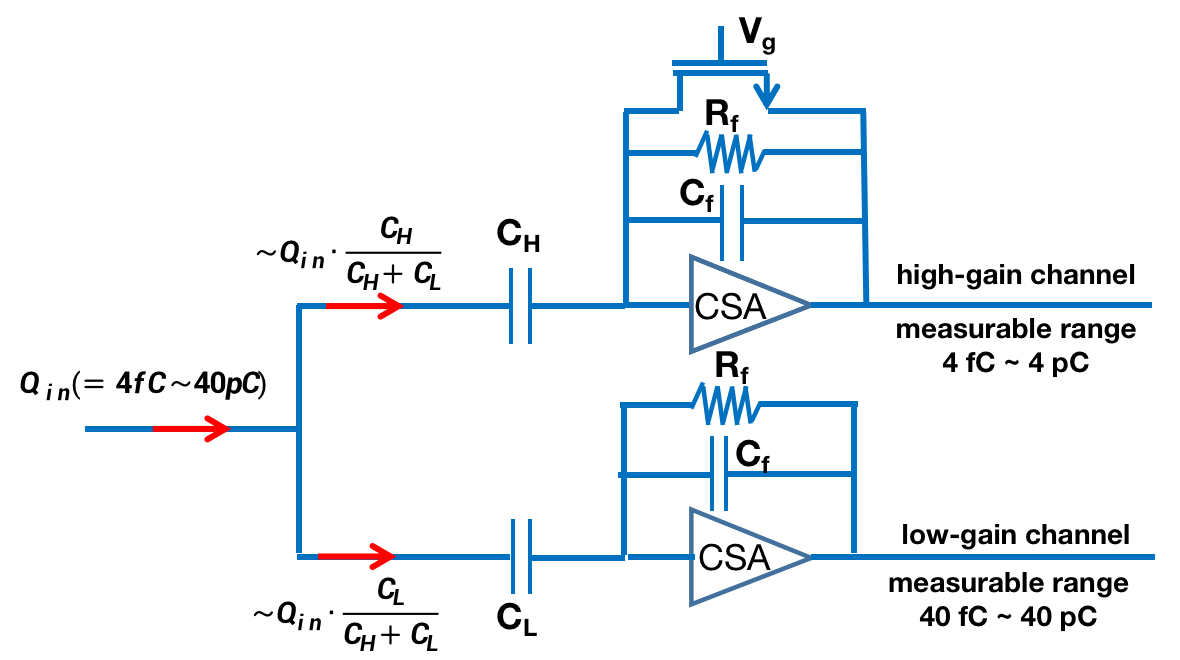}
 		\caption{\label{fig:circuit_diagram} The circuit scheme of the dual-gain charge-sensitive preamplifier (DGCSP) with saturation prevention circuit.}
 	\end{center}
 \end{figure}

This was achieved by custom electronics consisting of a new dual-gain charge-sensitive preamplifiers ASIC, called DGCSP (Fig.~\ref{fig:circuit_diagram}), that provides inputs to the ASIC chips called HINP16 version 3. The DGCSP chips were developed in RIKEN, Japan and the final boards were made in ATOMKI, Hungary. Some results of the tests with the first prototypes of the DGCSP done by the RIKEN team before the final version of the ASIC described in this paper were presented in Refs.\cite{Kurokawa2014,Takuma2012}. 
The second ASIC in the pulse-processing logic, developed by the Washington University and Southern Illinois University groups \cite{Engel2007}, has internal preamplifiers which can be bypassed, to directly access logic and linear pulse processing electronics. The logic branch has a constant-fraction discriminator (CFD) and the linear has two shapers (operating in parallel), each with a peak detect and hold circuit.

\begin{figure}[h!]
 	\begin{center}
 		\includegraphics[width=\columnwidth]{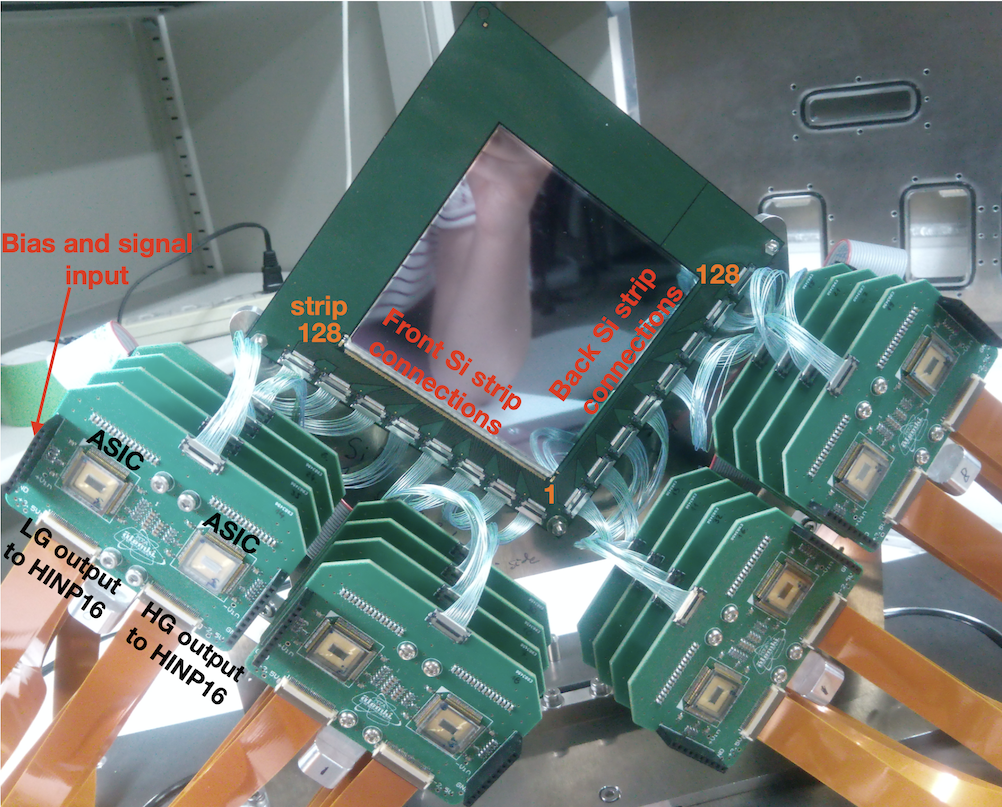}
 		\caption{\label{fig:setup} One pair of silicon detectors rotated at 45$^\circ$ and mounted on the support frame equipped with the necessary dual-gain preamplifiers (DGCSP).}
 	\end{center}
 \end{figure}

In total, 32 DGCSP boards are used to process 1024 signal channels (4 $\times$ 128 strips $\times$ 2 outputs); each silicon detector was serviced by 8 DGCSP boards. These boards are assembled into a compact stack-like structure and mounted as shown in Fig. \ref{fig:setup}.  Rotation of the silicon sensors by 45$^\circ$, allows for the preamplifier boards to be mounted close to the detectors and be operated inside the vacuum chamber. The DGCSP amplifier can provide a dynamic range of about 10$^{4}$ due to the dual-channel, dual-gain, dual-output design. This dynamic range comes at the cost of doubling the number of required down-stream channels, amplifying the need for an ASIC for the second stage provided by the HINP16.

The DGCSP ASIC was fabricated using a 0.5 {\textmu}m CMOS technology. As schematically illustrated in Fig.~\ref{fig:circuit_diagram}, the input charge from a silicon strip is divided asymmetrically in proportion to the ratio of the input impedances and, as such, is sensitive to the ratio between the external coupling capacitors: C$_{H}$ and C$_{L}$. To make sure that the capacitive division stage works correctly, the open-loop gain factor of the amplifier should be as large as possible. This ensures that the input impedance depends on the coupling  capacitor.
 
The feedback capacitors C$_{f}$ (= 1.8 pF) were added to stabilize the gain of the preamplifier and the feedback resistances R$_{f}$ (= 20 M$\Upomega$) were connected in parallel to C$_{f}$ with the purpose of preventing the pile-up by resetting the amplifier. Table~\ref{table2} shows the values of the coupling capacitors.
In addition to using the capacitance division method, it was necessary to add an anti-saturation circuit to the high-gain branch. In essence, when saturation occurs on the high-gain side, the input impedance becomes too large and the charge would flow to the low-gain side, which causes a reduction of the dynamic range.
To prevent this, a FET (field-effect transistor) was introduced in the feedback stage of the high-gain branch to allow the charge compensation at the input of the amplifier by making use of its charge sensitivity.  In this way, the gate bias voltage V$_{g}$ is adjusted depending on the saturation voltage of the amplifier. 

\begin{table}[h!]
\centering
\caption{Detection limits of the new DGCSP}\label{table2}
\resizebox{\columnwidth}{!}{\begin{tabular}{ c c c } 
\hline
  & High gain branch  & Low gain branch\\
\hline
Coupling capacitance    & C$_{H}$ = 5.6 nF  & C$_{L}$ = 0.56 nF  \\
Measurable range    & 4.0 fC -- 4.0 pC & 40 fC -- 40 pC  \\
Energy limits    & 90 keV -- 90 MeV  & 900 keV -- 900 MeV  \\
\hline
\end{tabular}}
\end{table}

With this newly developed dual-gain preamplifier, small (min. ~100 keV) and large (up to 900 MeV) energy loss signals are selectively processed in the high-gain and low-gain branches of the circuit and a 10$^{4}$ high dynamic range is realized.

A single front-end printed circuit board (PCB) carries two DGCSP chips. DGCSP serves 8 silicon detector strips. Thus eight low-gain signals and eight high-gain signals are outputted from each DGCSP chip. The high-gain (HG) and low-gain (LG) signals are separated at the PCB level (labelled with LG and HG in Fig. \ref{fig:setup}) and sent to different connectors of the PCB. The low-gain signals from two PCBs are merged and the 32 signals are sent to the processing circuit HINP16 chip boards each of which has two of the 16-ch HINP16 chips. A similar collection is made for the high-gain outputs.

Each dual-gain preamplifier board has an auxiliary 10-pins connector which distributes a test input pulse signal, the bias voltage for the chips ($\pm$ 3 V) and the bias for the silicon detectors (- 90 V). 

 The HINP16 ASIC  was used to assure the necessary triggering, shaping and amplification functions for the signal processing. As mentioned above, each HINP chip board consists of 2 HINP16 ASICs, i.e. services 32 channels. The HINP16 chip has an internal CSA with two gain modes: high-gain with 15 mV/MeV of 0.4 mV/fC and low-gain with 3 mV/MeV or 0.08 mV/fC, or these can be bypassed and used with an external amplifier. This work uses the latter option. The output of the CSA, or the external input, is split to feed two branches: one for energy and one for timing. The timing output results from charging a capacitor with a constant current source from the channel CFD to a user supplied common stop. To process the signals from the four silicon detectors, 32 HINP boards were used, 16 chip boards for each gain mode of the DGCSP.

\begin{figure}[h!]
 	\begin{center}
 		\includegraphics[width=\columnwidth]{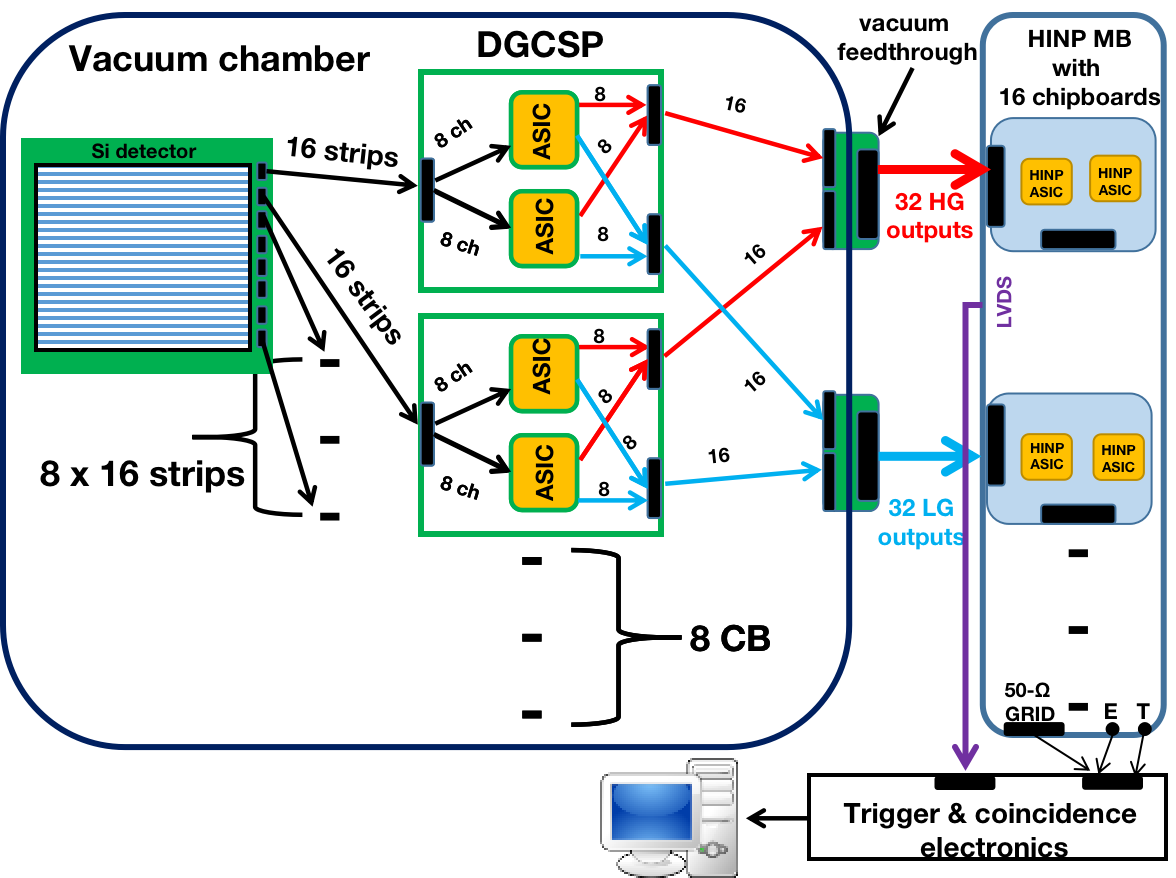}
 		\caption{\label{fig:scheme_new}  The electronics chain to process the signals from one silicon detector. In total 8 dual-gain preamplifiers boards (CB) are used for one silicon layer.}
 	\end{center}
 \end{figure}

To read and control the HINP chip boards, 2 HINP motherboards (MB) were used, each housing 16 HINP chip boards and thus 512 channels were serviced by each of the two MBs.
The software of the HINP system allows the user to enable/disable each input channel (corresponding to one detector strip) of the HINP16 chip from the discriminator mask, to set individual threshold for each chip channel, to select the polarity of the CSA input signal, to set a global gain mode for each HINP board and to inspect the CSA and Shaper signals. Also the motherboard can distribute a test input pulse signal to all boards, accordingly with their polarity set. The signal processing chain is presented in Fig.~\ref{fig:scheme_new}. The red arrows represent the HG signals and the blue the LG signals. The numbers above the arrows are the number of input/output signals transmitted through the electronics. With analog energy (E) and time (T) signals are sent from the MB to the ADC via a twisted-pair digital audio cables with Lemo 2-contact connectors. The digital HINP control signals, for both slow download of options and serially sequencing the analog changes off of the chip, are sent to and from the motherboard using a differential twisted-pair LVDS cable (colored with purple at the middle of the MB). To inspect the logical hit OR, CSA, CFD and multiplicity signals at the HINP chips, a 34-pin grid-connector is mounted on the MB and the signals are outputted on a 50 $\Upomega$ ribbon cable.

The HINP electronics was operated outside the vacuum and, to prevent chip overheating, the system was water cooled to 19$^\circ$C. Since the DGCSP boards were mounted in the same vacuum chamber close to the silicon detectors, to avoid worsening the energy resolution of the silicon detectors due to the heat dissipated by the dual-gain preamplifiers, the vacuum chamber itself was cooled to 14$^\circ$C.

\section{Performance}
\subsection{Pilot experiments at HIMAC facility}

The first tests of the newly constructed silicon tracker system were carried out at the National Institutes for Quantum Science and Technology (QST) in Japan using the HIMAC (Heavy Ion Medical Accelerator in Chiba) facility, under the H244 experiment, with beams of protons and various heavy ions (like $^{12}$C, $^{84}$Kr and $^{132}$Xe) at different energies. A limited number of prototype DGCSP chips and silicon detectors were used in the tests. The goal of these tests was to test the dynamic range of the DGCSP. 

\begin{figure}[ht!]
 	\begin{center}
		\includegraphics[width=\columnwidth]{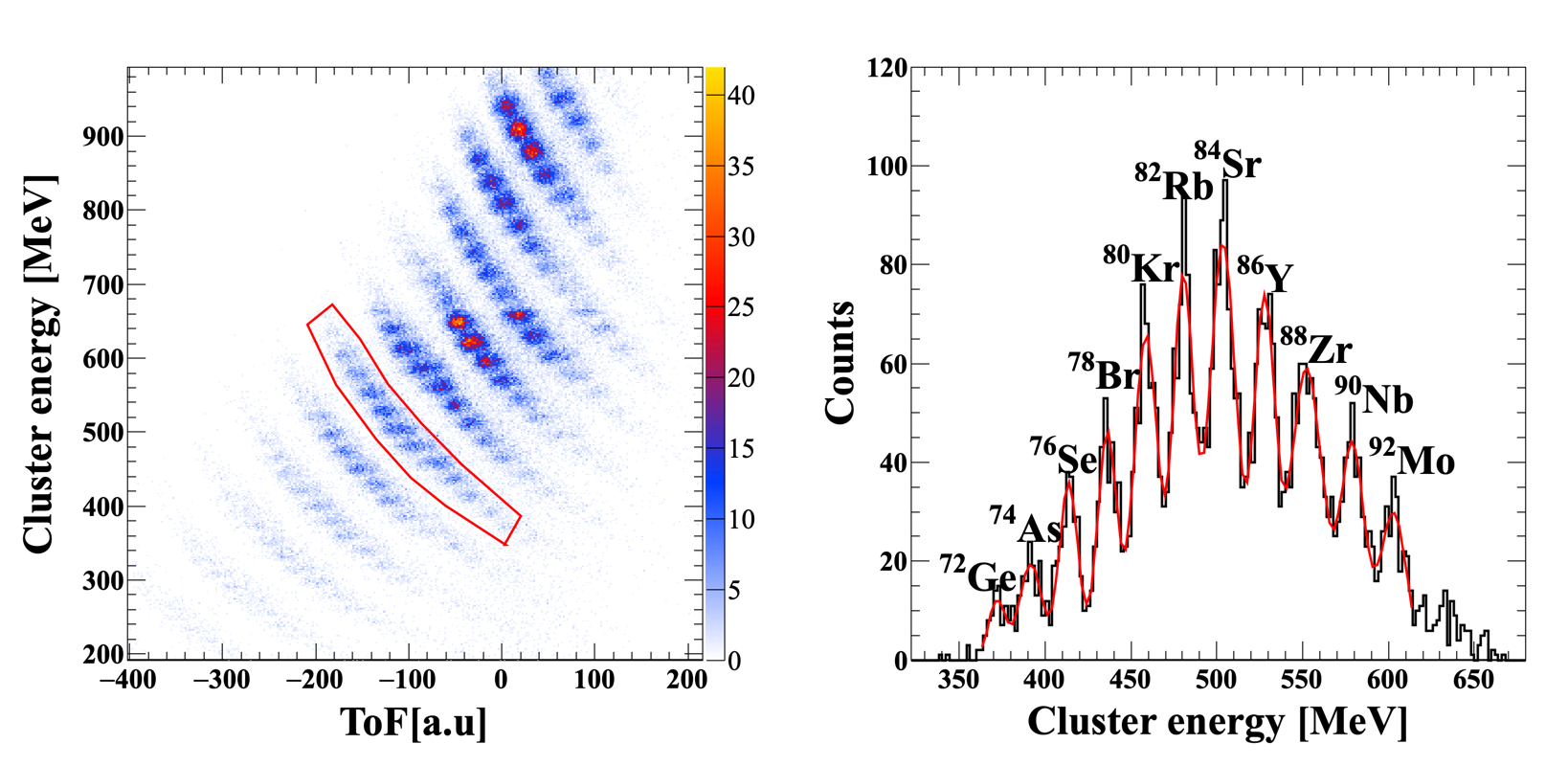}
 		\caption{\label{fig:PID_LG}  The experimental results for the DGCSP low-gain channel obtained during the HIMAC test experiment. In the left figure the 2D PID of the cocktail beam is shown and with red graphical cut the A/Q = 2.19 nuclei are selected. The right figure results when the data in the gate indicated by the red box is projected on the energy axis.}
 	\end{center}
\end{figure}

Figure~\ref{fig:PID_LG} shows the performance of the low-gain output of the DGCSP ASIC coupled to a silicon detector which was irradiated by a secondary cocktail beam of heavy ions produced in the fragmentation of primary $^{132}$Xe beam at 200 AMeV. The energy deposited in the silicon layer is marked with $\Delta$E and represents a cluster sum - integrated LG signals from a few adjacent strips fired by heavy ions. ToF is the time of flight of the ions through a fragment separator. The $\Delta$E resolution is about 1$\%$ over the wide dynamic range covered by the LG branch (up to 1 GeV). Gating (red box) and projecting on the energy axis allows for a 1D particle identification (PID) plot to be made, Fig. \ref{fig:PID_LG} right side.

\begin{figure}[ht!]
 	\begin{center}
		\includegraphics[width=\columnwidth]{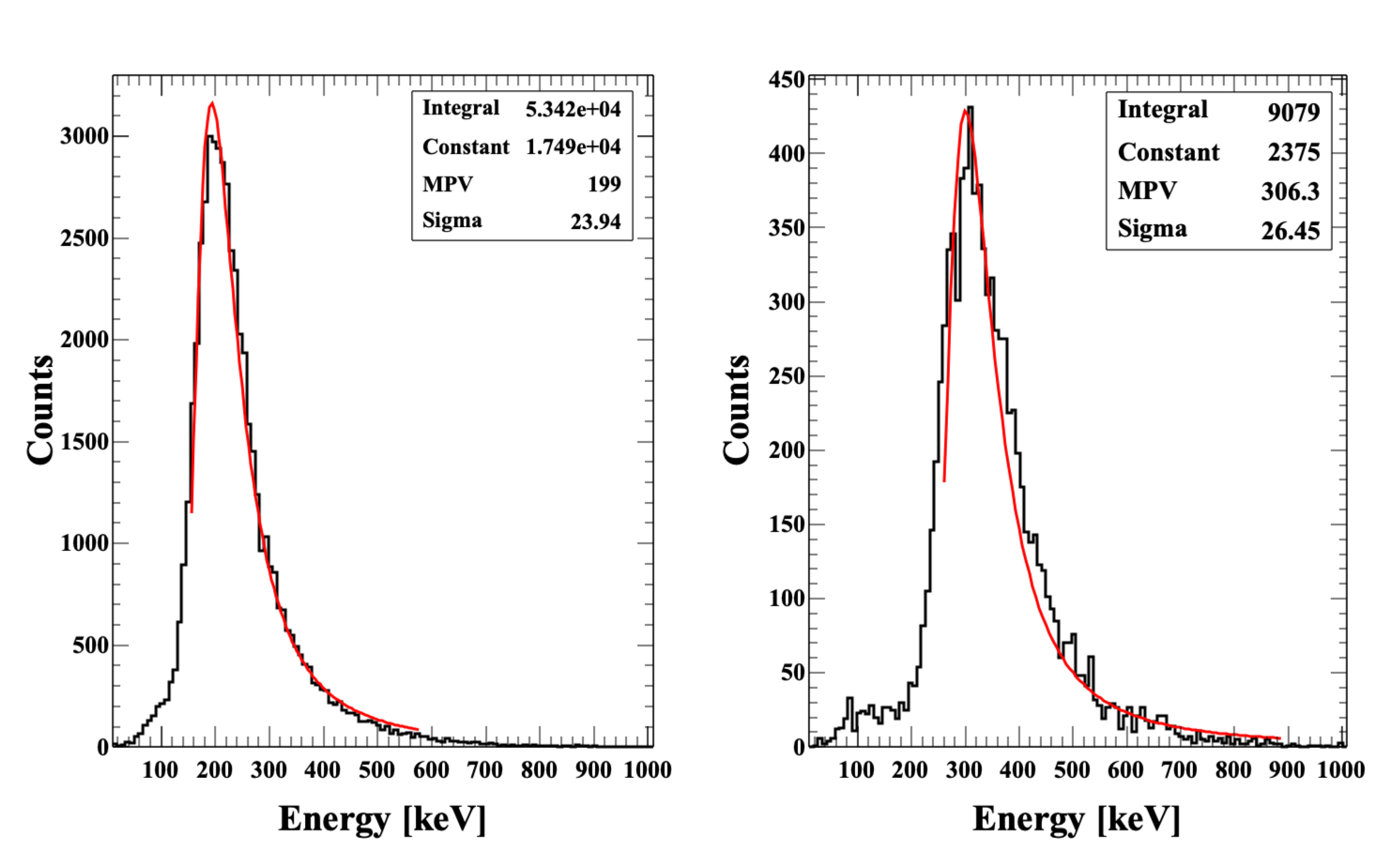}
 		\caption{\label{fig:proton_himac}  Proton signals measured with the HG channel during the HIMAC test experiment. In the left (right) figure are for protons with 230 MeV (150 MeV).}
 	\end{center}
\end{figure}

The $\Delta$E response of the HG readout for proton beams at two different energies is shown in Fig. \ref{fig:proton_himac}. The proton incident energies were selected to be similar to the ones expected in the two breakup experiments at SAMURAI. For the 230 MeV (150 MeV) protons, the deposited energy in the silicon layer is about 200 keV (300 keV).  These tests confirmed that the required dynamic range and low-energy detection threshold of about 100 keV was achieved by this electronics.

\subsection{SAMURAI experiments}

The GLAST silicon detectors with the signal processing described in this paper was foreseen to be used in breakup studies of proton-rich nuclei in experiments with the SAMURAI magnetic spectrometer at RIBF, in RIKEN (Japan). So far the system was used in two experiments: NP1412-SAMURAI29R1 (''Inclusive and exclusive breakup of $^{9}$C in nuclear and Coulomb fields'') and in NP1406-SAMURAI24 (''Investigation of proton-unbound states in neutron-deficient isotopes $^{66}$Se and $^{58}$Zn''). We will report here the performances of the detection system achieved during the $^{9}$C breakup study experiment\cite{rikenrep2019,nn2018}. For this experiment only the HG branch of the DGCSP ASICs was used as it alone had sufficient dynamic range to integrate the signals induced by all the reaction products.

\begin{figure}[h!]
 	\begin{center}
 		\includegraphics[width=\columnwidth]{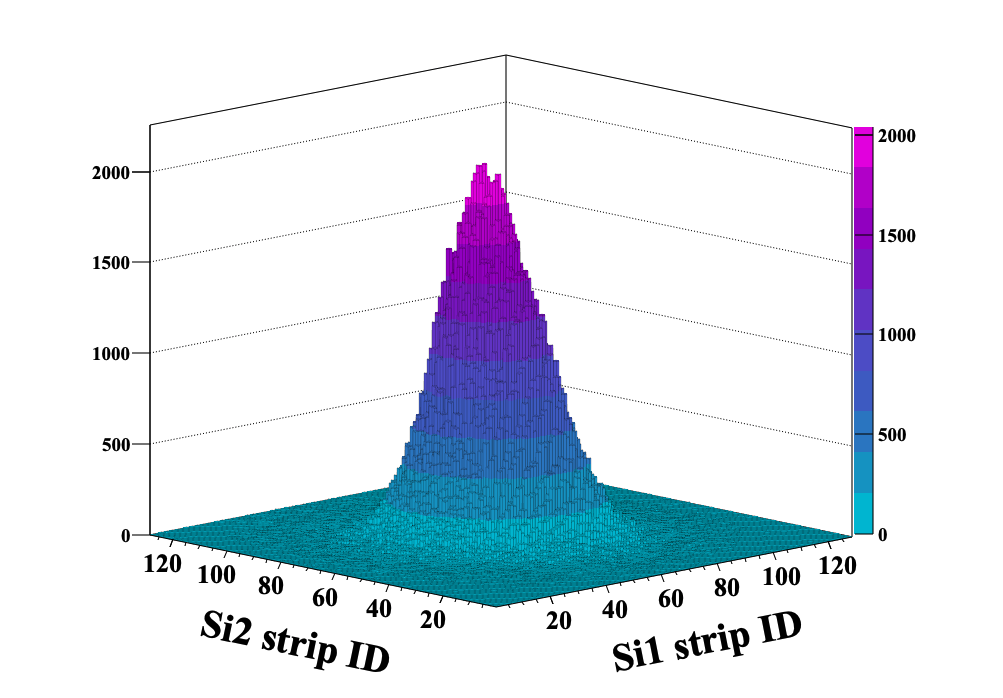}
 		\caption{\label{fig:beam_spot}  Beam spot plot in the first pair of silicon detectors.}
 	\end{center}
 \end{figure}

In Fig. \ref{fig:beam_spot} the beam spot for the $^{9}$C is shown and it proves that the active area of the GLAST silicon detectors is large enough to ensure geometrical acceptance for tracking. 

\begin{figure}[h!]
 	\begin{center}
 		\includegraphics[width=\columnwidth]{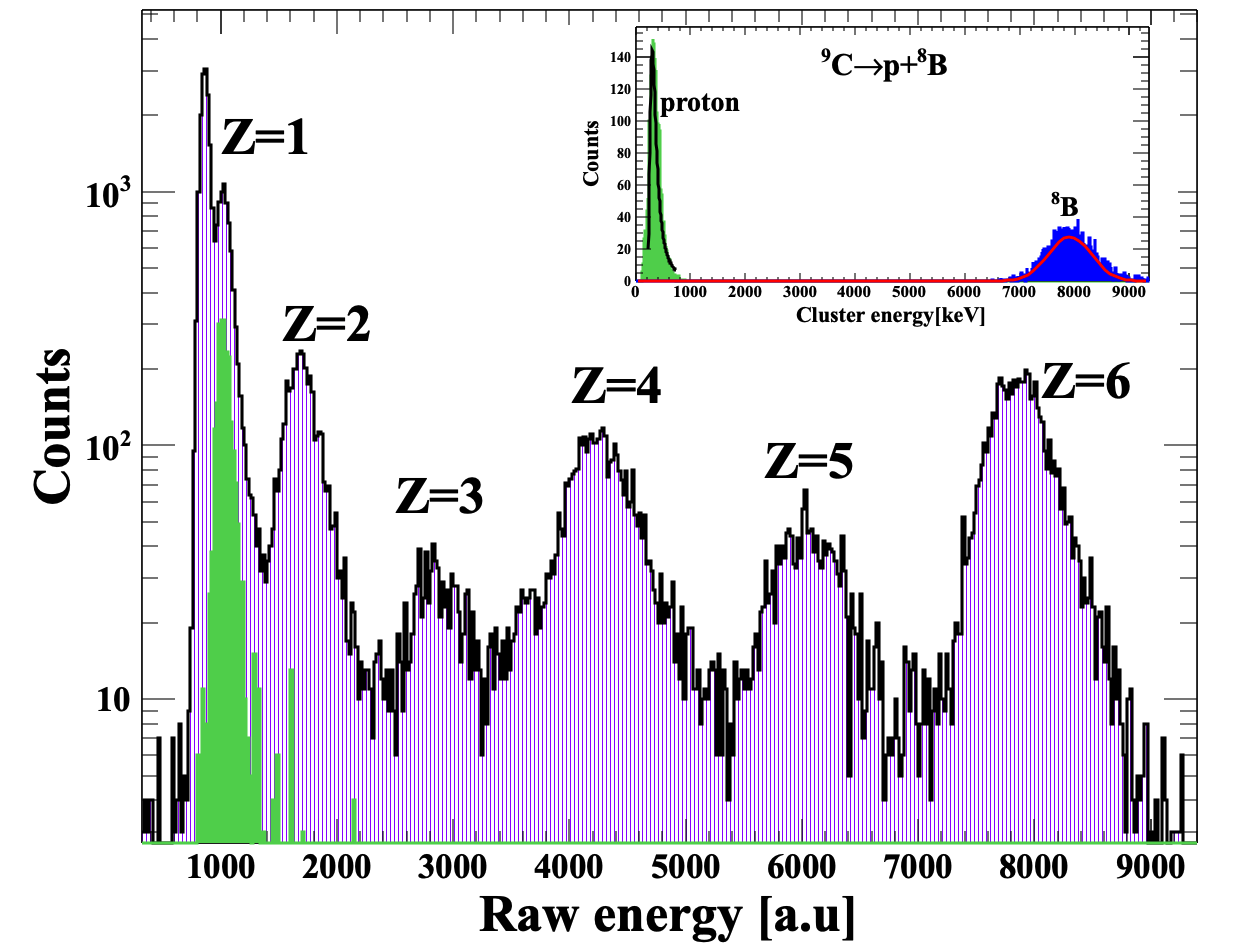}
 		\caption{\label{fig:charges}  $^{9}$C fragmentation products after passing through Pb target - black histograms, proton peak - green histogram. In the inset, the one-proton breakup channel of $^{9}$C is shown.}
 	\end{center}
 \end{figure}

By using the silicon tracking detectors in conjunction with the standard SAMURAI detection systems it was possible to reconstruct the necessary PID, right after the target. As can be seen in Fig. \ref{fig:charges}, the silicon system can well distinguish between the charges of the nuclei induced by the beam interaction with the target. At the beginning of the experiment, a defocused proton beam was sent through the silicon detectors for calibration and the corresponding energy loss spectrum is indicated by the green peak.  The inset figure shows the 1-proton removal channel from $^{9}$C, measuring in coincidence the proton and the $^{8}$B core. These data use the first silicon layer and require the produced particles to be confirmed by the other SAMURAI detectors.

\begin{figure}[ht!]
 	\centering
	\begin{subfigure}[b]{0.23\textwidth}
		\centering
 		\includegraphics[width=\textwidth]{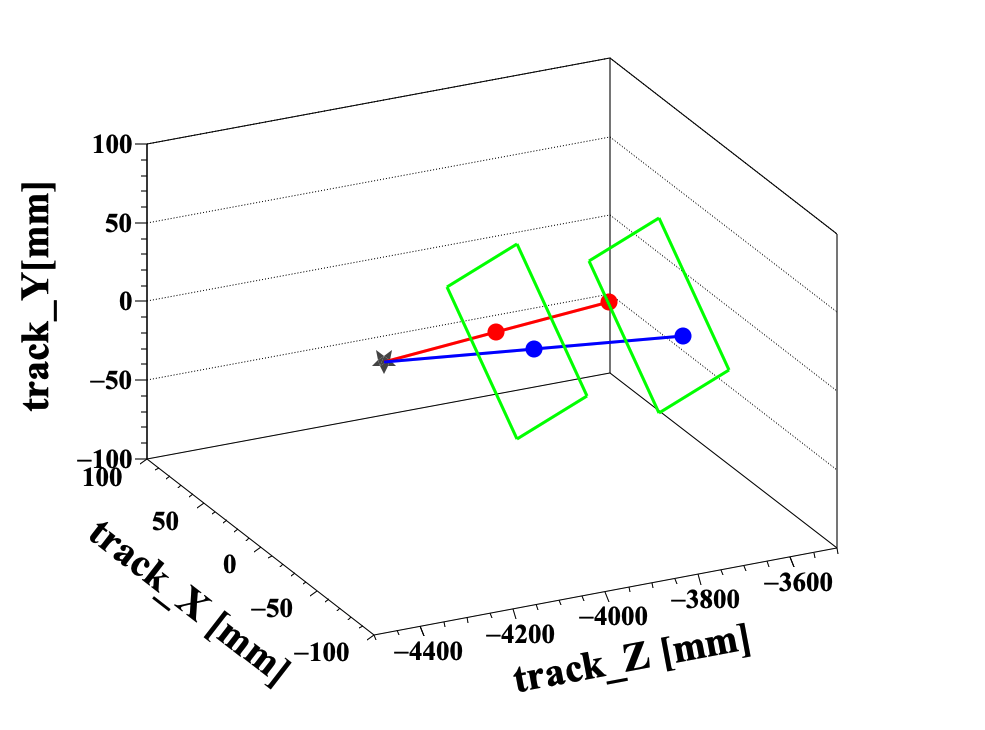}
		\caption{}
		\label{fig:tracks}
		\end{subfigure}
	\begin{subfigure}[b]{0.23\textwidth}
		\centering
 		\includegraphics[width=\textwidth]{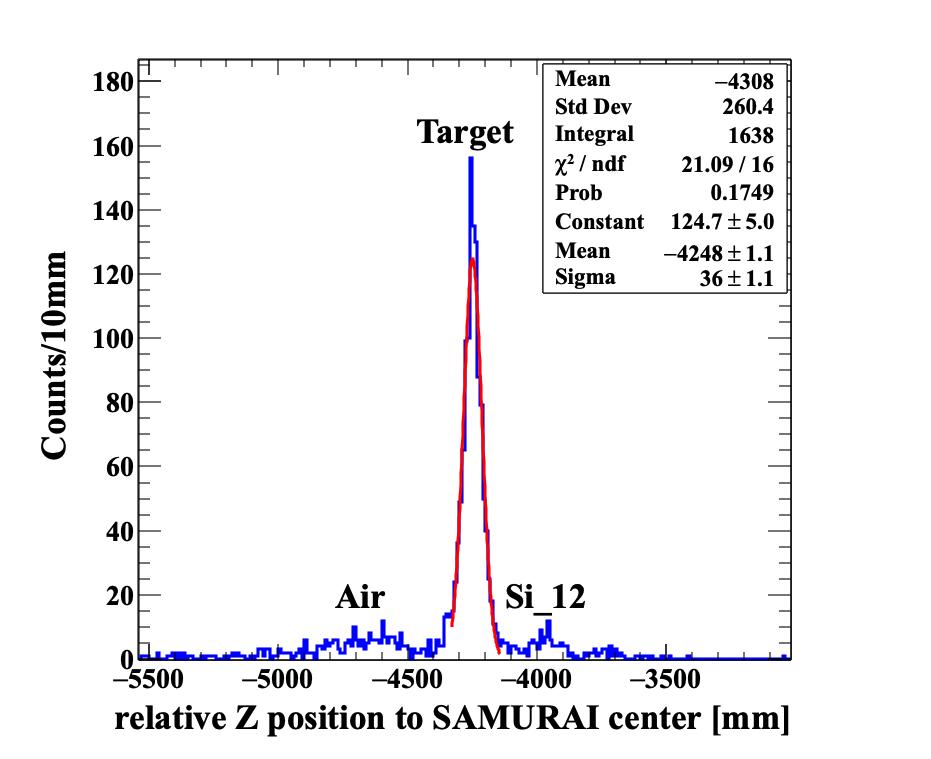}
		\caption{}
		\label{fig:Zposition}
		\end{subfigure}
\caption{\label{fig:vertex_reconstruction}  Vertex point reconstruction relative to the center of the SAMURAI magnetic spectrometer: (\subref{fig:tracks}) Position measurement of a $^{9}$C breakup event with the silicon detectors and (\subref{fig:Zposition}) The breakup vertex reconstruction around the target area.}
 \end{figure}

Figure~\ref{fig:vertex_reconstruction}(\subref{fig:tracks}) shows an example of the proton and $^{8}$B trajectories after the breakup, reconstructed using the position information measured with the silicon detectors. Figure~\ref{fig:vertex_reconstruction} (\subref{fig:Zposition}) shows a spectrum of vertex positions in the beam axis resulting from the reconstruction. This vertex reconstruction helps to remove the unwanted events from the beam interaction with non-target material in the beam path, e.g., the air before the target and the first pair of the silicon detectors that can themselves produce nuclear reactions.

\begin{figure}[h!]
 	\begin{center}
 		\includegraphics[width=0.9\columnwidth]{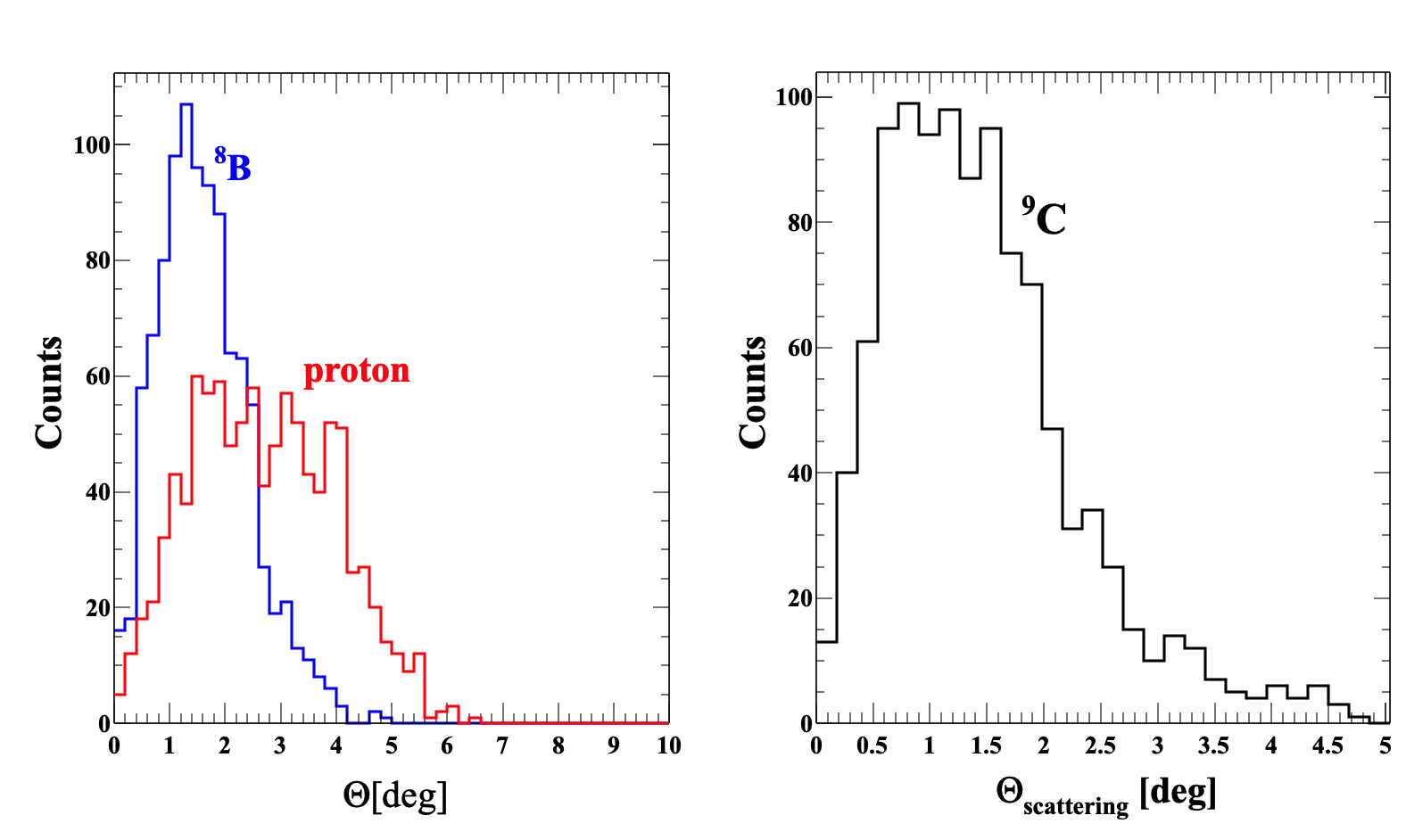}
 		\caption{\label{fig:angular_reconstruction}  Angular distributions of the projectile and of the emerging particles. Left figure: the proton emission angle (red) and the $^{8}$B emission angle (blue) from the breakup of $^{9}$C on a Pb target.  Right figure: the $^{9}C^{\ast}$ scattering angle on the same Pb target.}
 	\end{center}
 \end{figure}

Using the drift chambers placed before the target, the actual angle of each incident $^{9}$C in the beam was determined. Together with the position information in all 4 silicon detectors, we determined the emission angles of the products ($^{8}$B nuclei and protons) after break-up. Fig.~\ref{fig:angular_reconstruction} left shows the angular distributions. With the momentum information from the SAMURAI spectrometer we finally obtained the scattering angle of excited $^{9}$C just before breakup or of the centre-of-mass of the p - $^{8}$B system (Fig.~\ref{fig:angular_reconstruction} right).
 
\section{Summary}

A silicon tracker for use in front of the SAMURAI spectrometer at the RIBF facility of the Nishina Center of RIKEN, Wako, Japan was designed, built and tested. The tracker consists of two pairs of single-sided microstrip detectors that determine the (x,y) position at two locations along the beamline after the target. The system provides particle identification from protons up to Z $\sim$ 50 heavy fragments with energies 100 -- 350 AMeV.  GLAST-type 325 {\textmu}m thick detectors were used. The inherent problems stemming from the large granularity, the required very large dynamic range and counting rate were handled using two matched ASICs: a dual-gain preamplifier DGCSP and the HINP16 pulse processing system. 

Combining the HINP16 ASICs with the newly developed dual-gain preampliﬁers, produces $2 \times 4 \times 128$ = 1024 channels. (Lo/Hi gain $\times$ 4 Si each with 128 channels.) This system yields a very large dynamic range of $\sim$ 10$^{4}$. It can work in a self-triggering mode or in slave mode where it requires an external trigger and can tolerate very large input capacitance. The large dynamic range, provided by the dual-channel preampliﬁer, makes it possible to measure a wide range of nuclear charges. Parts of the system were tested with beams from the HIMAC facility in Chiba, Japan, to show that it can measure energy losses from 100 keV (protons) to 600 -- 900 MeV (for heavy fragments up to Z $=$ 50), which makes it appropriate for the studies with radioactive ion beams at intermediate energies.

The whole system was used and characterized in two RIBF experiments (NP1412-SAMURAI29 and NP1406-SAMURAI24), using the SAMURAI magnetic spectrometer. The silicon detection system was successfully used to track protons and heavy fragments simultaneously in the breakup of the proton-rich nuclei, (like $^{9}$C and $^{66}$Se) to reconstruct the reaction vertex, the emerging angles of the particles, the momentum distributions of the protons and the relative energy spectra. It dramatically extends the research opportunities with the SAMURAI spectrometer especially for systems with two or more charge particles, which had been too difficult with the standard detectors equipped in SAMURAI.

\begin{acknowledgements}
We acknowledge the participation of Mathew McCleskey, Brian T. Roeder and Hiroyuki Murakami in the initial phases of the project.
We acknowledge the RIKEN Nishina Center accelerator staff for providing the stable beams and the accelerator staff from the QST (Chiba) for their stable operation of the accelerators of the HIMAC facility.
This work was supported by the U.S. DOE Office of Science under Grant DE-SC0004971, by the Romanian Ministry of Research and Innovation Bucharest under grant No. PN-III-P4-ID-PCE-2016--0743 and the PNIII-P5-P5.2 No. 02/FAIR-RO program with the NAIRIB and NAFRO projects, by the Kakenhi Grant-in-Aid for Young Scientists B No. 16K17719 and by the Hungarian Academy of Sciences under Grants No. NN114454-NKFIH and NN128072-NKFIH.
AIS and DT acknowledge the support of the IPA program of RIKEN for their stay in Wako, Japan. CAB acknowledges support by the U.S. DOE grant DE-FG02-08ER41533.
\end{acknowledgements}

\bibliography{mybibfile.bib}   % name your BibTeX data base

%\begin{thebibliography}{}
% Format for Journal Reference Author, Article title, Journal, Volume, page numbers (year)
%\bibitem{RefB} Author, Book title, page numbers. Publisher, place (year)
% etc
%\end{thebibliography}

\end{document}